\documentclass{article}

\usepackage{arxiv}

\usepackage[utf8]{inputenc} 
\usepackage[T1]{fontenc}      
\usepackage{hyperref}       	
\usepackage{url}            
\usepackage{booktabs}       
\usepackage{amsfonts}       
\usepackage{amsmath}
\usepackage{nicefrac}       
\usepackage{microtype}      
\usepackage{cleveref}       
\usepackage{graphicx}
\usepackage{natbib}
\usepackage{doi}
\usepackage{svg}
\usepackage{algorithm}
\usepackage[noend]{algpseudocode}
\usepackage{subfig}
\usepackage{tcolorbox}
\usepackage{siunitx}
\usepackage{lipsum}
\usepackage{xparse}
\usepackage{ifthen}
\tcbuselibrary{theorems}
\newtcolorbox{llm}[1]{colback=green!5,colframe=green!35!black,fonttitle=\bfseries, title={#1}}
\newtcolorbox{prompt}[1]{colback=blue!5,colframe=blue!35!black,fonttitle=\bfseries, title={#1}}

\def\defmysvgtype{svg}

\NewDocumentCommand{\includemysvg}{O{} m}{%
	\ifthenelse{\equal{\defmysvgtype}{svg}}%
	{\includesvg[#1]{#2}}%
	{\includegraphics[#1]{svg-inkscape/#2_svg-tex.pdf}}%
}

\makeatletter
\def\BState{\State\hskip-\ALG@thistlm}
\algnewcommand\algorithmicforeach{\textbf{for each}}
\algdef{S}[FOR]{ForEach}[1]{\algorithmicforeach\ #1\ \algorithmicdo}
\renewcommand{\Function}[2]{\csname ALG@cmd@\ALG@L @Function\endcsname{#1}{#2}\def\jayden@currentfunction{#1}}
\newcommand{\funclabel}[1]{\@bsphack\protected@write\@auxout{}{\string\newlabel{#1}{{\jayden@currentfunction}{\thepage}}}\@esphack}
\makeatother

\title{QuST-LLM: Integrating Large Language Models for Comprehensive Spatial Transcriptomics Analysis}

\author{ \href{https://orcid.org/0000-0002-3837-8135}{\includegraphics[scale=0.06]{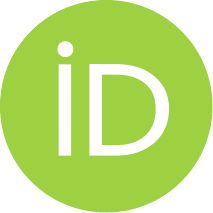}\hspace{1mm}Chao Hui Huang}\\
	Pfizer Inc.\\
	La Jolla, CA 92101 \\
}

\renewcommand{\shorttitle}{QuST-LLM: Integrating LLMs for Comprehensive ST Analysis}

\hypersetup{
pdftitle={QuST-LLM: Integrating LLMs for Comprehensive ST Analysis},
pdfsubject={q-bio.NC, q-bio.QM},
pdfauthor={Chao Hui Huang},
pdfkeywords={Large language model, spatial transcriptomics, whole slide images, QuPath extension},
}

\begin{document}
\maketitle

\begin{abstract}

In this paper, we introduce QuST-LLM, an innovative extension of QuPath that utilizes the capabilities of large language models (LLMs) to analyze and interpret spatial transcriptomics (ST) data. In addition to simplifying the intricate and high-dimensional nature of ST data by offering a comprehensive workflow that includes data loading, region selection, gene expression analysis, and functional annotation, QuST-LLM employs LLMs to transform complex ST data into understandable and detailed biological narratives based on gene ontology annotations, thereby significantly improving the interpretability of ST data. Consequently, users can interact with their own ST data using natural language. Hence, QuST-LLM provides researchers with a potent functionality to unravel the spatial and functional complexities of tissues, fostering novel insights and advancements in biomedical research.  QuST-LLM is a part of QuST project. The source code is hosted on GitHub and documentation is available at \url{https://github.com/huangch/qust}.

\end{abstract}


\section{Introduction}

Spatial transcriptomics (ST) (\cite{Stahl:2016}) has emerged as a transformative technology in the field of genomics, enabling the high-resolution mapping of gene expression across tissue sections. This spatial context is crucial for understanding the cellular architecture and functional organization of tissues. Traditional transcriptomics, which averages gene expression across entire tissues or cell populations, often obscures the spatial heterogeneity and cell-to-cell variability that are fundamental to tissue function and disease progression (\cite{Janesick:2023, NatureMethods:2021}), as well as its capabilities of bridging single-cell data to the corresponding pathological image (\cite{Bergenstrahle:2022,Huang:2023b}).

With the advancements in ST technologies, there has been a surge of interest in large language models (LLMs) (\cite{Devlin:2018, Brown:2020}). Researchers have recognized the potential of LLMs and have begun leveraging their capabilities to expedite computational biology research, particularly in the field of ST.

For example, Bioinformatics Copilot 1.0, introduced by \cite{Wang:2024}, is a tool powered by a large language model. This tool enables intuitive data analysis through a natural language interface, without requiring programming skills, and supports cross-platform functionality. This tool is with a great potential to accelerate advancements in the biomedical sciences by expediting the data analysis workflow. Also, \cite{Choi:2024} developed a framework called CELLama that uses a language model to transform cell data into "sentences" that capture gene expressions and metadata, allowing for universal cellular data embedding and analysis. CELLama has the potential to revolutionize cellular analysis by enabling cell typing and analysis of spatial contexts without the need for manual reference data selection or dataset-specific workflows.

There are existing methods that may or may not utilize LLMs as natural language interpreters in the task of decoding ST data, but these methods do involve technologies relevant to LLMs. For example, \cite{Luo:2024} proposed StereoMM, a machine learning toolchain that integrates gene expression, histological images, and spatial location data, providing an advanced analysis platform for effectively utilizing multimodal and high-throughput data from spatially resolved omics technologies. \cite{Ji:2024} proposed SpaCCC, a method for inferring spatially resolved cell-cell communications in ST data using a fine-tuned single-cell language model and a functional gene interaction network. SpaCCC embeds ligand and receptor genes into a unified latent space and identifies likely interacting pairs based on their proximity in this space. Lastly, \cite{Cui:2024} proposed scGPT, a foundational model for single-cell biology that effectively distills biological insights and can be optimized for superior performance in tasks like cell type annotation, multi-omic integration, and gene network inference.

\begin{figure}[tb]
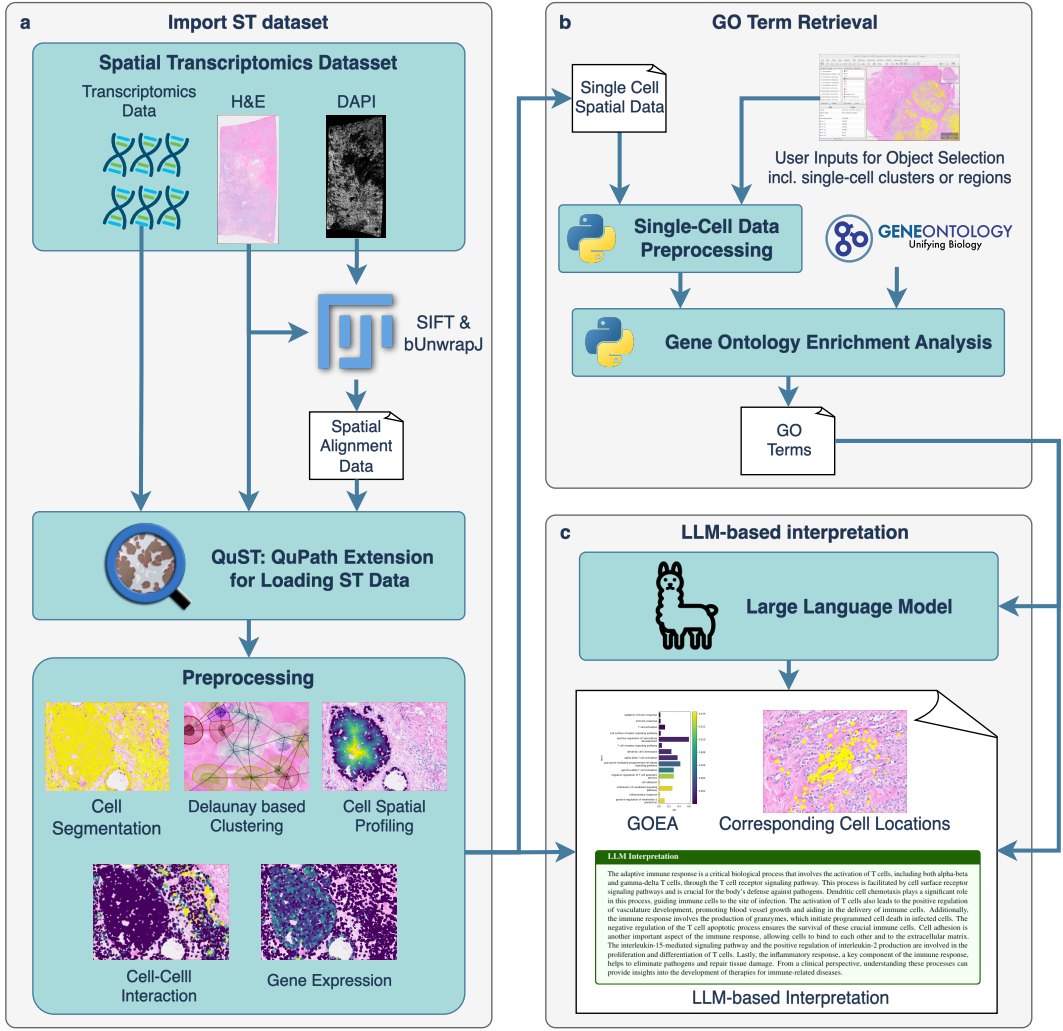

	\centering
	\includemysvg[width = 0.85\linewidth]{diagram1}
	\caption{The QuST-LLM workflow for forward analysis includes the following steps: (a), users begin by importing ST data into QuPath using QuST. This step may require additional spatial alignment data, which can be obtained via FIJI if the user is working with a 10x Xenium dataset (see text for more details). Once the ST data is successfully loaded, users can perform analysis and visualization using QuPath and QuST. (b), QuST-LLM takes the objects selected by the user, including single-cell clusters or regions, performs a series of single-cell data preprocessing steps and then obtains a list of GO terms based on GOEA. (c), the spatial data and GO terms are integrated as biological evidence, which can be interpreted using an LLM service. The final outcomes is presented to the users.}
	\label{fig:diagram1}
\end{figure}

Despite its potential, the interpretation of ST data presents significant challenges due to the complexity and volume of the information it generates. Advanced computational tools are required to manage, analyze, and interpret these data effectively. Current approaches often rely on a combination of bioinformatics tools to preprocess and analyze the data, but they fall short in providing comprehensive biological insights. To address this need, we present QuST-LLM, an extension of QuPath that enables computational biologists to explore spatial biological problems while providing visualization and analysis capabilities for whole slide image (WSI) analysis. 

\begin{figure}[tb]
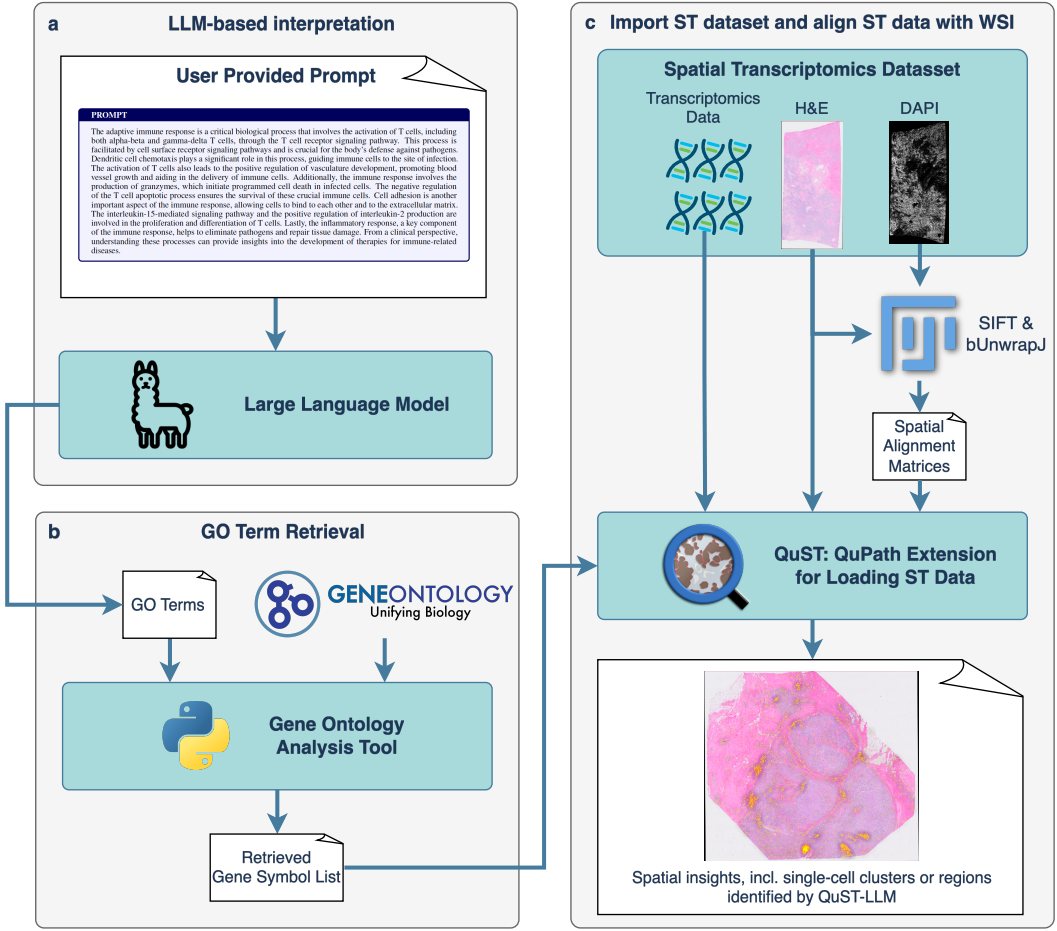

	\centering
	\includemysvg[width = 0.85\linewidth]{diagram2}
	\caption{The QuST-LLM workflow for backward analysis includes the following steps: (a), users begin by providing languages describing the required biological evidences. A LLM service is then interpreting the inputs and obtains the the key terms which may be used to isolate the sub-graph of the GO. (b), QuST-LLM identifies the key genes by using GOEA based on the obtained GO terms. (c), given the ST data which has been loaded into QuST, the users can then identify the cells which may highly relevant to the sentences provided by the users.}
	\label{fig:diagram2}
\end{figure}

QuST-LLM utilites a LLM as the backbone. LLMs, such as GPT-4, have demonstrated remarkable abilities in natural language processing, including the interpretation and generation of human-like text based on vast amounts of data. By leveraging LLMs and gene ontology (GO) (\cite{Ashburner:2000}), a knowledge graph containing , QuST-LLM can translate complex biological annotations into accessible and comprehensive explanations, significantly enhancing the interpretability of ST data.

\section{Methods}
\label{sec:methods}

QuST-LLM relies on two major components: QuPath (\cite{Bankhead:2017}) and QuST (\cite{Huang:2024}). QuPath is an open-source software platform widely used for bioimage analysis, offering powerful tools for visualizing and analyzing high-resolution tissue images. QuST is a powerful extension for QuPath that seamlessly integrates whole slide image (WSI) and ST analysis, providing enhanced capabilities for spatial biology research. It enables the visualization of spatial gene expression data alongside histopathological images, allowing researchers to explore the molecular landscape of tissues at an unprecedented resolution. In this section, we will introduce the analyzing tools and use cases available in QuST.

QuST-LLM facilitates a seamless workflow from data acquisition to biological interpretation. First, users load ST data into QuPath, select regions of interest (ROIs). Next, QuST-LLM takes care of the following analyzing tasks automatically, including  profiling gene expression within these regions using SCANPY (\cite{Wolf:2018}), identifying the corresponding genes, and performing GO enrichment analysis (GOEA) using GOATOOLS (\cite{Klopfenstein:2018}). Finally, the GO terms are interpreted using LLMs, providing detailed explanations of the biological significance of the selected cells.

One of the challenges in analyzing ST data is aligning it with WSI, as different image modalities are involved. QuST addresses this challenge by offering various data loading approaches for different ST data formats, including 10x Visium, 10x Xenium, NanoString CosMX, and more. Each format requires a specific approach for proper alignment.

Once the data is loaded, QuST provides a range of visualization and analysis tools. Researchers can visualize gene expression patterns in specific tissue regions, overlay multiple data layers, and adjust visualization parameters to highlight different aspects of the data. This flexibility enables detailed exploration of the spatial relationships between gene expression and tissue morphology. By leveraging QuST's capabilities, researchers can gain valuable insights into the spatial biology of tissues, advancing our understanding of cellular organization and function.

There are two scenarios of using QuST-LLM: forward and backward analyses. The two scenarios will be introduced in the following sub-sections.

\subsection{Forward Analysis: Interpreting Spatial Data using LLM}

In the forward analysis (see Figure~\ref{fig:diagram1}), the user begins by loading spatial transcriptomics (ST) data into a software tool called QuST. The specific steps involved in this process may vary depending on the chosen data modality. However, in general, the following procedure is typically required:
\begin{enumerate}
	\item Load the high-resolution whole slide image (WSI) into QuPath.
	\item Perform necessary analysis on the WSI, including cell segmentation to identify each individual cells.
	\item Compute and load spatial correspondence between the WSI and ST data, including the Scale-Invariant Feature Transform (SIFT) (\cite{Lowe:2004}) affine matrix and the B-spline basis function coefficients (\cite{Sorzano:2005}), to QuST.
	\item Select the targeting single-cell clusters or regions.
	\item Use QuST-LLM to interpret the targeting single-cell clusters or regions. 
\end{enumerate}

\begin{figure}[t]
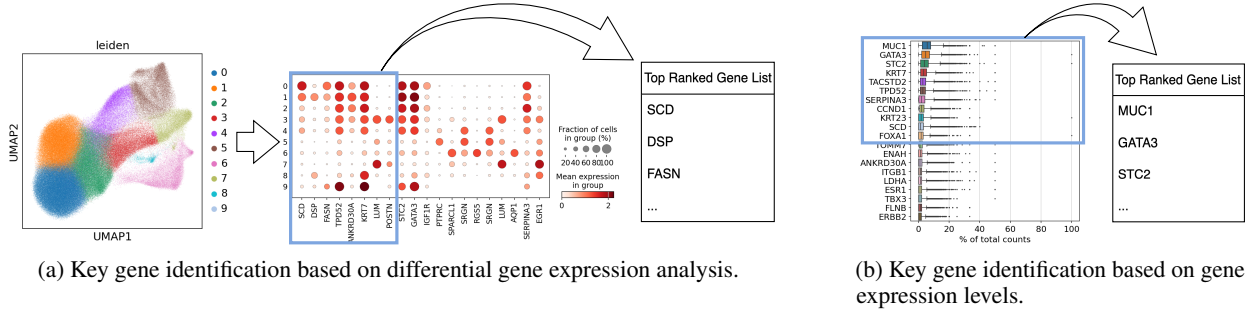

	\centering%
	\subfloat[Key gene identification based on differential gene expression analysis.]{\label{fig:keygenes:a}%
		\includemysvg[height = 3.25cm]{comparative-genes}%
	}%
	\hfill%
	\subfloat[Key gene identification based on gene expression levels.]{\label{fig:keygenes:b}%
		\includemysvg[height = 3.25cm]{high-ranking-genes}%
	}%
	\caption{Two approaches for obtaining key genes.}%
	\label{fig:keygenes}
\end{figure}

There are two options available for ST data interpretation using QuST-LLM:
\paragraph{Interpretation based on high ranking key genes.} In this option, QuST-LLM interprets the ST data based on key genes which are the genes playing critical roles in the user-specific single-cell clusters or regions (see Figure~\ref{fig:keygenes:a}). In the case if the differential gene expression analysis is insufficient, the gene expression level can be used to measure the ranking of genes (see Figure~\ref{fig:keygenes:b}).

\begin{figure}[t]
	\centering%
	\subfloat[The chosen single cells labeled by yellow marker.]{\includegraphics[height=7cm]{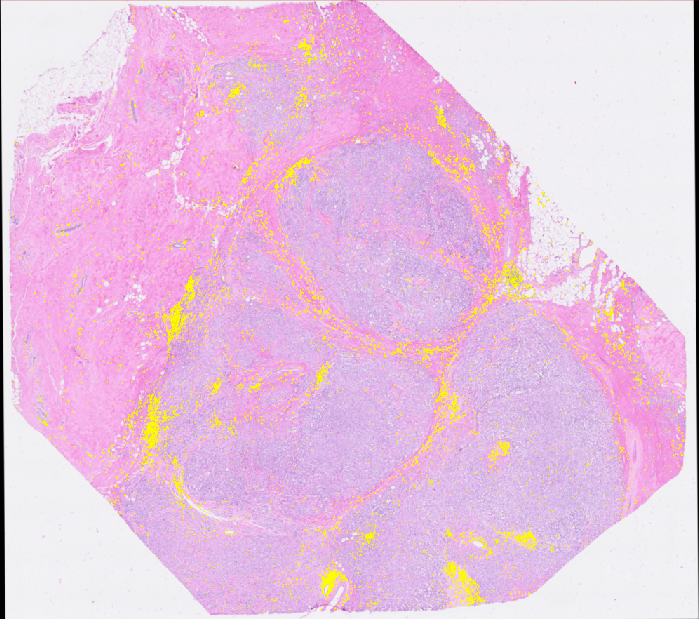}}%
	\hfill%
	\subfloat[The corresponding GOEA.]{\includegraphics[height=7cm]{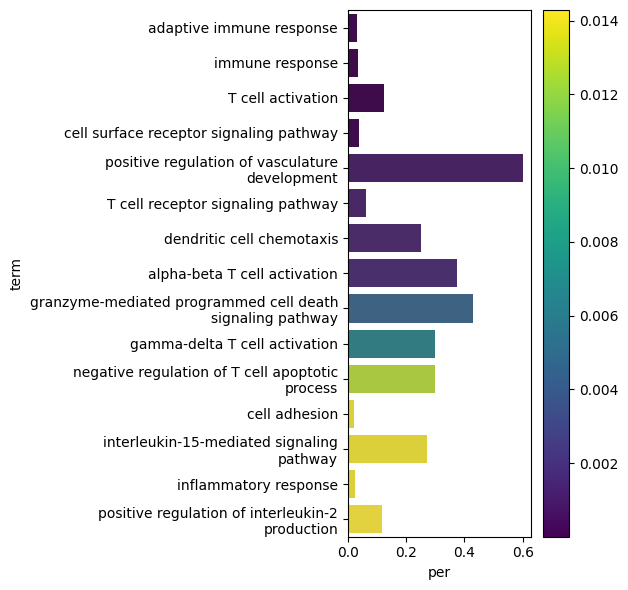}}%
	\\%
	\subfloat[The interpretation generated by LLM.]{\label{fig:result1:c}%
		\begin{llm}{LLM Interpretation}%
			{\footnotesize The adaptive immune response is a critical biological process that involves the activation of T cells, including both alpha-beta and gamma-delta T cells, through the T cell receptor signaling pathway. This process is facilitated by cell surface receptor signaling pathways and is crucial for the body's defense against pathogens. Dendritic cell chemotaxis plays a significant role in this process, guiding immune cells to the site of infection. The activation of T cells also leads to the positive regulation of vasculature development, promoting blood vessel growth and aiding in the delivery of immune cells. Additionally, the immune response involves the production of granzymes, which initiate programmed cell death in infected cells. The negative regulation of the T cell apoptotic process ensures the survival of these crucial immune cells. Cell adhesion is another important aspect of the immune response, allowing cells to bind to each other and to the extracellular matrix. The interleukin-15-mediated signaling pathway and the positive regulation of interleukin-2 production are involved in the proliferation and differentiation of T cells. Lastly, the inflammatory response, a key component of the immune response, helps to eliminate pathogens and repair tissue damage. From a clinical perspective, understanding these processes can provide insights into the development of therapies for immune-related diseases.}%
	\end{llm}}%
	\caption{LLM interpretation of high ranking genes based on the selected immuno-cell clusters. (a) The provided whole slide image (WSI) with highlighted single-cell clusters indicated by yellow spots. (b) The results of GOEA, with the x-axis representing the ratio of relevant genes and relevant GO terms,  the y-axis showing the list of identified GO terms sorted by uncorrected p-values, and the heat map represents the corresponding p-value for each GO term. (c) The interpretation of the selected immuno-cell clusters as determined by LLM.}%
	\label{fig:result1}
\end{figure}

\paragraph{Interpretation based on comparative key genes.}
In this option, QuST-LLM interprets the ST data based on the differential expression of genes between two or more single-cell clusters or regions defined by the user. This means that the analysis focuses on genes that show significant differences in expression levels between the selected clusters or regions.

These options allow users to gain insights into the spatial distribution and patterns of gene expression within the tissue of interest using the QuST-LLM tool.

\subsection{Backward Analysis: Discovering Spatial Insights based on Human Languages using LLM}

In the backward analysis (see Figure~\ref{fig:diagram2}), the user starts from providing a description of the desired targeting single-cell clusters or regions. And then, QuST-LLM will identify these single-cell clusters or regions accordingly. The general procedure is as the following:
\begin{enumerate}
	\item Load the high-resolution whole slide image (WSI) into QuPath.
	\item Perform necessary analysis on the WSI, including cell segmentation to identify each individual cells.
	\item Compute and load spatial correspondence between the WSI and ST data, including the SIFT affine matrix and B-spline basis function coefficients, to QuST.
	\item User provides a content in human languages describing the required biological evidences. 
	\item Use QuST-LLM to identify the targeting single-cell clusters or regions. 
\end{enumerate}
The first three steps are identical to the procedure in the forward analysis as they are essential steps for loading WSI and ST data into QuST. This procedure generates a measurement per cell indicating the level of relevance to the provided description.

\begin{algorithm}[tbp]
	\caption{Algorithm to compute the relevance of the given ST data to the user-provided natural language descriptions.}\label{alg:asn}
	\begin{algorithmic}[1]
		\State {\textbf{definition:}}
		\State {$g$: a string representing a gene symbol defined in GO.}
		\State {$\mathcal{G}=\{g_1,g_2,\cdots\}$: a set of gene symbols available in the given ST dataset.}
		\State {$t$: a string representing a GO term ID.}
		\State {$x$: GO category, one of \textit{biological process (BP)}, \textit{molecular function (MF)} and/or \textit{cellular component (CC)}.}
		\State {$\mathcal{N}_g^{(x)}=\{t_1,t_2,\cdots\}$: a set as a dictionary that maps $g$ to  a set of associate GO terms based on the given $x$.}
		\State {$\mathcal{M}^{(x)}=\{\mathcal{N}_{g_1}^{(x)},\mathcal{N}_{g_2}^{(x)},\cdots\}$: a set of dictionaries $\mathcal{N}_g^{(x)}$ of the given GO category $x$.}
		\State {$w_g\in\mathbb{R},\forall g\in\mathcal{G}$: the weight for $g$ to be computed, }
		\State {$\mathcal{W}=(w_{g_1},w_{g_2},\cdots)$: a list of $w_g,\forall g\in\mathcal{G}$.}
		\State {$c$: the symbol representing a cell.}
		\Statex
		\State {\textbf{input:}}
		\State {$\mathcal{T}=\{t_{1}, t_{2}, \cdots\}$: the GO term list obtained via LLM based on the user-provided description in human language.}
		\State{$\mathcal{C}=\{c_1,c_2,\cdots\}$: list of all targeting cells.}
		\State {$\mathcal{K}^{(c)}=\{g_1^{(c)},g_2^{(c)},\cdots,g_k^{(c)}\}\subseteq\mathcal{G},\forall c\in\mathcal{C}$: top $k$ ranked genes for $c$.}
		\Statex
		\State{\textbf{procedure:}}
		\State{$\mathcal{W}\gets\mathbf{0}$}\Comment{initial $\mathcal{W}$.}
		\Statex
		\ForEach {$x\in\{\text{'BP'}, \text{'MF'}, \text{'CC'}\}$}\Comment{compute $w_g\in\mathcal{W},\forall g\in\mathcal{G}$  based on the given $\mathcal{T}$.}
			\ForEach {$g\in\mathcal{G}$}
				\If {$\mathcal{N}_g^{(x)}\in\mathcal{M}^{(x)}$}
					\State {$w_g\gets w_g+|\mathcal{N}_g^{(x)}\cap\mathcal{T}|$}
				\EndIf
			\EndFor
		\EndFor
		\Statex
				\State {$\mathcal{W}\gets\mathcal{W}/||\mathcal{W}||_1$}\Comment{normalize  $\mathcal{W}$.}
		\Statex
		\ForEach {$c\in\mathcal{C}$}\Comment{compute $r^{(c)}$ based on $\mathcal{W}$ and $\mathcal{K}^{(c)},\forall c\in\mathcal{C}$.}
			\State {$r^{(c)}\gets\sum_{g\in\mathcal{K}^{(c)}}w_g$}
		\EndFor
		\Statex
		\State {\textbf{output:}}
		\State {$r^{(c)}\in[0,1],\forall c\in\mathcal{C}$: the relevance between cell $c$ and the description provided by the user in human language.}
	\end{algorithmic}
\end{algorithm}

In the $4th$ step, the user selects specific single-cells or regions to target. QuST-LLM performs quality control (QC) and identifies key genes strongly associated with the chosen regions. In the 5th step, QuST-LLM translates the provided description into a list of Gene Ontology (GO) terms linked to a set of genes. Finally, QuST-LLM calculates the correlation between the genes from the identified GO terms and the high ranking genes  (either based on their expression levels directly, or based on the correlation levels to some gene sub-groups) in the selected single-cell clusters or regions. The resulting correlation values for each cell are stored in the measurement table in QuPath for further investigation. The pseudo code showing in Algorithm~\ref{alg:asn} shows how to compute the relevance between a cell and the description provided by the user in human language.
 
 \begin{figure}[t]
 	\centering
 	\subfloat[The chosen region labeled by yellow marker.]{\label{fig:result2a}\includegraphics[height=7.5cm]{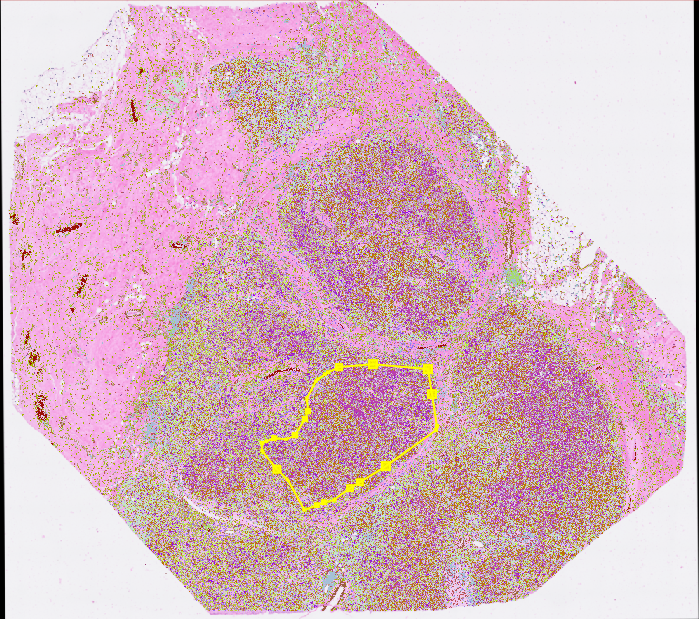}}%
 	\hfill%
 	\subfloat[The corresponding GOEA.]{\includegraphics[height=7.5cm]{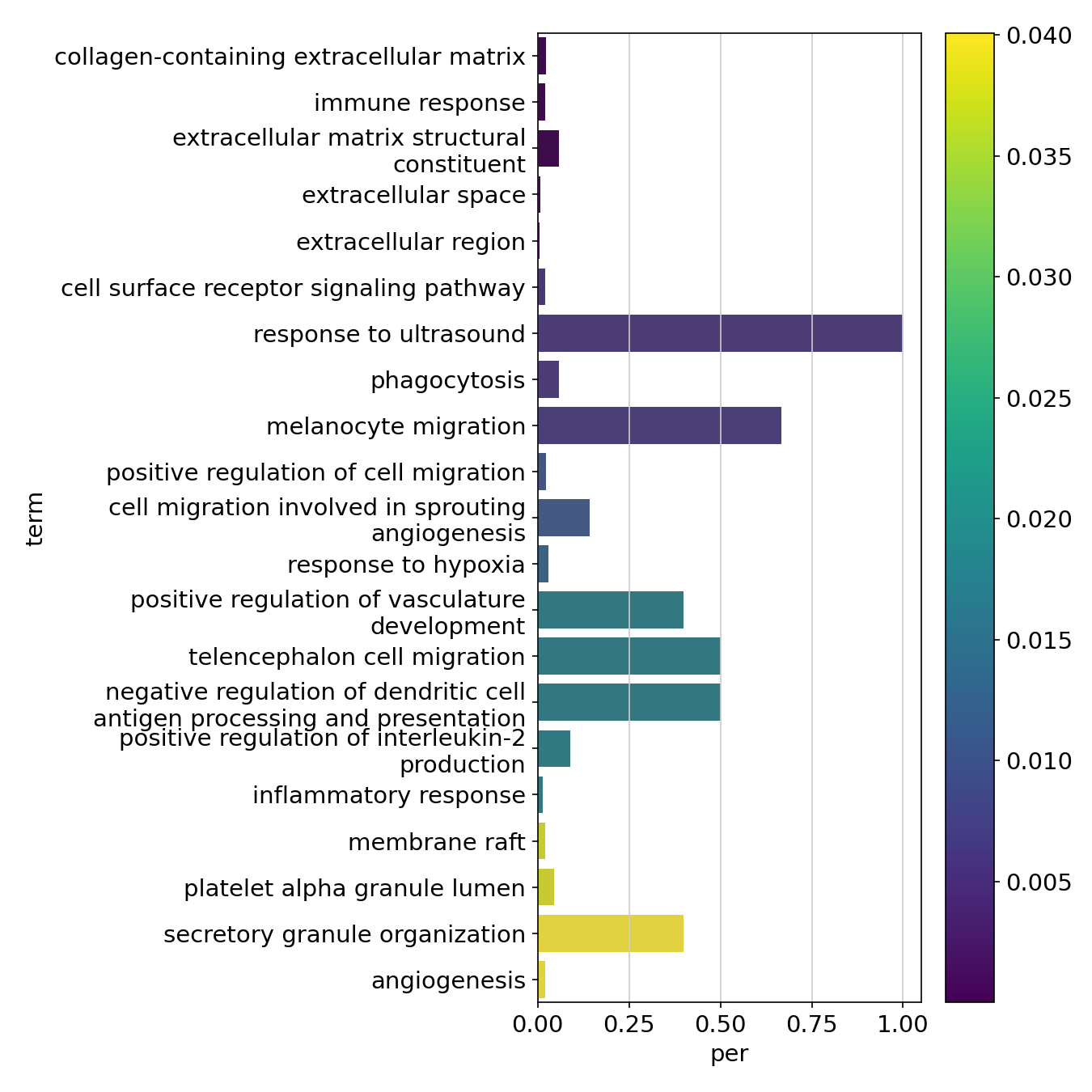}}%
 	\\%
 	\subfloat[The interpretation generated by LLM.]{
 		\begin{llm}{LLM Interpretation}
 			{\footnotesize The biological processes, cellular components, and molecular functions of certain cell groups interact in complex ways to maintain tissue structure and function, with implications for clinical perspectives. The immune response, particularly the adaptive immune response, plays a crucial role in regulating cell migration and monocyte chemotaxis, both of which are essential for tissue repair and inflammation. Calcium-mediated signaling is a key player in these processes, and it also influences the production of interleukin-4, a cytokine involved in immune responses. Interestingly, these processes can be modulated by ultrasound, suggesting potential therapeutic applications. 

			\hspace{15pt} The extracellular matrix, particularly the collagen-containing component, provides a structural framework for tissues and is involved in various biological processes such as angiogenesis and response to hypoxia. It also plays a role in the organization of secretory granules and the formation of extracellular exosomes, which are involved in cell-cell communication. 
 				
			\hspace{15pt} The process of angiogenesis, or the formation of new blood vessels, is crucial for tissue repair and regeneration. It involves cell migration and is influenced by factors such as hypoxia and the metabolic process of triglycerides. 
			
			\hspace{15pt} Lastly, the migration of melanocytes, cells that produce the pigment melanin, is a key process in skin pigmentation and can be influenced by various factors, including the extracellular matrix and immune responses. Understanding these interactions can provide insights into conditions such as vitiligo, where melanocyte migration is disrupted.}%
 	\end{llm}}
 	\caption{LLM interpretation of the selected epithelial/tumor-epithelial single-cell clusters based on high ranked gene expression. (a) The provided whole slide image (WSI) with highlighted epithelial/tumor-epithelial single-cell clusters indicated by yellow markers. (b) The result of GOEA, with the x-axis representing the ratio of relevant genes and relevant GO terms, and the y-axis showing the list of identified GO terms sorted by uncorrected p-values. The heat map represents the corresponding p-value for each GO term. (c) The interpretation generated by LLM.}
 	\label{fig:result2}
 \end{figure}
 
\section{Results}

In our experiment, we used the data\footnote{\url{https://www.10xgenomics.com/datasets/ffpe-human-breast-with-custom-add-on-panel-1-standard}} prepared by 10x~Genoimics that was generated for  demonstrating gene expression profiling for formalin-fixed paraffin-embedded (FFPE) human breast samples using the Xenium platform. The sample was obtained as \si{5\micro\meter} sections from resected tumor mass tissues of a  invasive lobular carcinoma obtained from from Avaden Biosciences Inc. The data was generated using 10x Xenium pre-designed panel along with an add-on panel of 100 custom genes. Additionally, the experiment was also conducted using the pre-designed Xenium Human Breast Gene Expression panel. The tissue preparation protocols followed were Xenium In Situ for FFPE - Tissue Preparation Guide and Xenium In Situ for FFPE Tissues – Deparaffinization \& Decrosslinking. Post-instrument processing was done according to the Xenium In Situ Gene Expression - Post-Xenium Analyzer H\&E Staining protocol.

For annotating regions, due to the fact that the raw H\&E image wasn't suitable for more refined exact sub-cellular resolution correspondence due to minute differences in the optics of the microscopy systems, the image was corrected using the methods described at the H\&E to Xenium DAPI Image Registration with FIJI Analysis Guide\footnote{\url{https://www.10xgenomics.com/analysis-guides/he-to-xenium-dapi-image-registration-with-fiji}}. As a result, the generated SIFT affine matrix and B-spline transformation matrix can be used in the QuST when necessary.

\subsection{Interpreting ST Data using LLM}

In this sub-section, we will present the results of LLM-based ST data analysis. The approaches of gene selection include: 1) high gene expression and 2) differential gene expression. The LLM used in this experiment was GPT-4, provided by OpenAI Inc.\footnote{\url{https://openai.com/}}. 

\subsubsection{ST Analysis using LLM-based Approach for High Ranking Gene Expression}
\label{sec:llm4heg}

Figure~\ref{fig:result1} illustrates the QuST-LLM interpretation of high ranking gene based on the selected immuno-cell clusters. In the experiment, the genes were selected based on the expression level. 

In this experiment, the prompt was:
\begin{center}
	\parbox{0.9\textwidth}{
		\texttt{``Write a paragraph of an integrative and comprehensive summary for the below given key gene ontological terms which are identified by an analysis of a single-cell dataset. Focusing on clinical meaning and structural biology. \\\\\{GO\_TERM:$1$, GO\_TERM:$2$, $\cdots$, GO\_TERM:$n$\},''}}
\end{center}
where the GO terms were obtained using GOATOOL based on the identified genes of interests in each of the experiments.

In this experiment. QuST-LLM shored detailed insights into the activation of T cells, including alpha-beta and gamma-delta T cells, through the T cell receptor signaling pathway. Additionally, QuST-LLM explores dendritic cell chemotaxis, which guides immune cells to the site of infection, and highlights the role of T cell activation in promoting vasculature development for efficient immune cell delivery. Furthermore, QuST-LLM discusses the involvement of Granzymes in inducing programmed cell death in infected cells, and emphasizes the importance of negative regulation of T cell apoptosis for their survival. Cell adhesion, interleukin signaling, and the inflammatory response are also identified as crucial components of the immune response.

Figure~\ref{fig:result2} shows the result of en experiment that performed LLM-based interpretation of tumor regions. Given the fact that the chosen regions include various cell types (see Figure~\ref{fig:result2a}), we used differential gene expression analysis to identify the key genes for this experiment. 

In this experiment, the prompt was:
\begin{center}%
	\parbox{0.9\textwidth}{%
		\texttt{``Write a paragraph of a summary for the given content. In the below given content, each row represent some biological processes, cellular components and/or molecular functions of a certain cell group. The summary focuses on the interaction (if any) among these groups. The paragraph is integrative and comprehensive, focus on tissue biological structure and clinical perspectives.
			\\\\\{GO\_TERM:$1$, GO\_TERM:$2$, $\cdots$, GO\_TERM:$m$\},\\\\%
			\{GO\_TERM:$3$, GO\_TERM:$4$, $\cdots$, GO\_TERM:$n$\},\\\\$\cdots$,''}}%
\end{center}%
where each row of GO terms represented a category identified using differential gene expression analysis.

The  interpretation that QuST-LLM generated suggested the fact that the nature of the chosen regions, including their biological processes, cellular components, and molecular functions interact in complex ways to maintain tissue structure and function, offering potential clinical insights. The the chosen regions, the immune response, especially the adaptive immune response, and calcium-mediated signaling are crucial in regulating cell migration and monocyte chemotaxis for tissue repair and inflammation. Further, the extracellular matrix, particularly its collagen component, provides a structural framework for tissues and contributes to various biological processes including angiogenesis and response to hypoxia. In addition, the extracellular matrix also influences the organization of secretory granules and formation of extracellular exosomes, which are key to cell-cell communication. Understanding the process of angiogenesis and the migration of melanocytes, influenced by factors like the extracellular matrix and immune responses, can provide insights into conditions such as vitiligo.

\subsubsection{ST Analysis  using LLM-based Approach for Differential Gene Expression}
\label{sec:llm4deg}

\begin{figure}[t]
	\centering%
	\subfloat[The chosen targeting region (labeled by yellow marker) and the contrast region (labeled by cyan marker).]{\includegraphics[height=7cm]{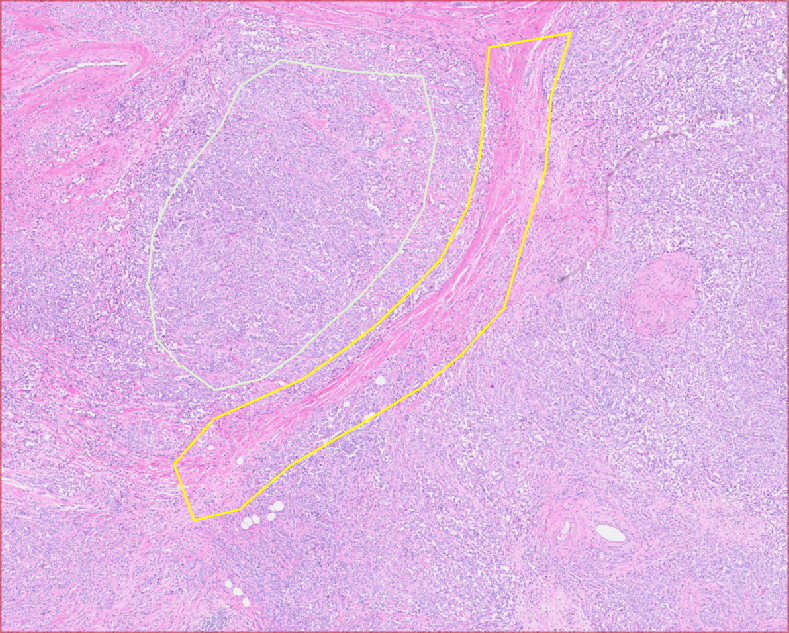}}%
	\hfill%
	\subfloat[The corresponding GOEA.]{\includegraphics[height=7cm]{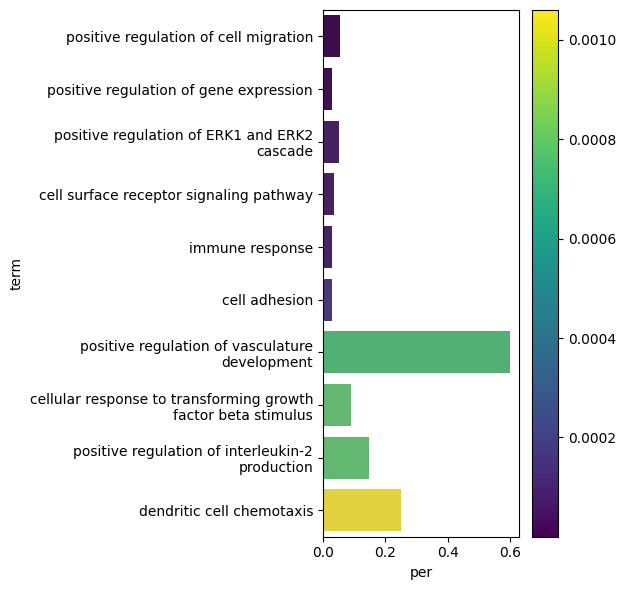}}%
	\\%
	\subfloat[The interpretation generated by LLM.]{%
		\begin{llm}{LLM Interpretation}{\footnotesize The positive regulation of cell migration, gene expression, and ERK1 and ERK2 cascade are crucial processes in the biological structure of tissues, playing significant roles in cellular communication, growth, and differentiation. These processes are often mediated by cell surface receptor signaling pathways, which are integral to the immune response and cell adhesion. The positive regulation of vasculature development is another key aspect, contributing to the formation and maintenance of blood vessels within tissues. The cellular response to transforming growth factor beta stimulus is also noteworthy, as it influences various cellular functions, including proliferation, differentiation, and apoptosis. Furthermore, the positive regulation of interleukin-2 production is essential for immune responses, as it promotes the growth and differentiation of T cells, B cells, and natural killer cells. Lastly, dendritic cell chemotaxis, the directed movement of dendritic cells in response to certain chemicals, plays a pivotal role in the immune response, as these cells act as messengers between the innate and adaptive immune systems. From a clinical perspective, understanding these processes can provide insights into the pathogenesis of various diseases and potentially guide the development of therapeutic strategies.}%
	\end{llm}}%
	\caption{LLM interpretation of differential gene expression based on the selected tumor-stromal region, in contrast to a tumor region. (a) The provided whole slide image (WSI) with highlighted tumor-stromal region indicated by yellow regions and  tumor region  indicated by cyan marker. (b) The results of gene ontology enrichment analysis (GOEA), with the x-axis representing the ratio of relevant genes and relevant GO terms, and the y-axis showing the list of identified GO terms sorted by uncorrected p-values. The heat map represents the corresponding p-value for each GO term. (c) The LLM interpretation of the selected tumor-stromal region, in contrast to a tumor region.}%
	\label{fig:result3}%
\end{figure}%

Figure~\ref{fig:result3} shows an interpretation based on differential gene expression between a tumor-stormal region and a tumor-epithelial region. QUST-LLM highlighted important processes such as cell migration, gene expression, and the ERK1 and ERK2 cascade in tissue biology. These processes are regulated by cell surface receptor signaling pathways, which are crucial for immune response and cell adhesion. Vasculature development is also a key aspect, contributing to the formation and maintenance of blood vessels. QuST-LLM also provide additional knowledge, \textit{e.g.,} the cellular response to transforming growth factor beta stimulus influences various cellular functions, and the positive regulation of interleukin-2 production is essential for immune responses, dendritic cell chemotaxis, which enables their movement and communication between the innate and adaptive immune systems, further plays a pivotal role, \textit{etc.} Thus, the experiment result suggested that QuST-LLM is able to build a shortcut for understanding these processes, eventually  provides valuable insights into disease pathogenesis and offers potential guidance for the development of therapeutic strategies.

\subsection{Discovering Spatial Insights based on Human Languages using LLM}
\label{sec:llm4ci}

\begin{figure}[t]
	\centering
	\subfloat[The prompt used in the experiment.]{%
	\begin{prompt}{PROMPT}%
		{\footnotesize The adaptive immune response is a critical biological process that involves the activation of T cells, including both alpha-beta and gamma-delta T cells, through the T cell receptor signaling pathway. This process is facilitated by cell surface receptor signaling pathways and is crucial for the body's defense against pathogens. Dendritic cell chemotaxis plays a significant role in this process, guiding immune cells to the site of infection. The activation of T cells also leads to the positive regulation of vasculature development, promoting blood vessel growth and aiding in the delivery of immune cells. Additionally, the immune response involves the production of granzymes, which initiate programmed cell death in infected cells. The negative regulation of the T cell apoptotic process ensures the survival of these crucial immune cells. Cell adhesion is another important aspect of the immune response, allowing cells to bind to each other and to the extracellular matrix. The interleukin-15-mediated signaling pathway and the positive regulation of interleukin-2 production are involved in the proliferation and differentiation of T cells. Lastly, the inflammatory response, a key component of the immune response, helps to eliminate pathogens and repair tissue damage. From a clinical perspective, understanding these processes can provide insights into the development of therapies for immune-related diseases.}%
\end{prompt}}%
	\\%
	\subfloat[The ground truth.]{\includegraphics[width=0.33\linewidth]{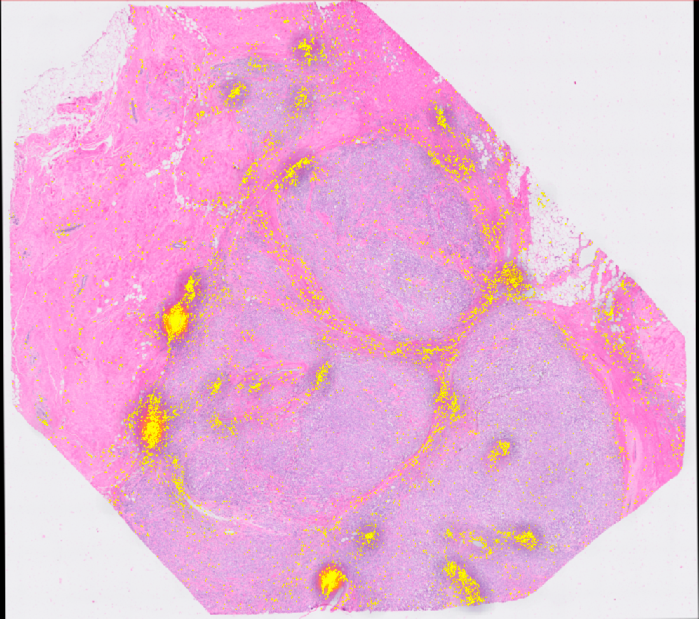}}%
	\hfill%
	\subfloat[The predicted outcome.]{\includegraphics[width=0.33\linewidth]{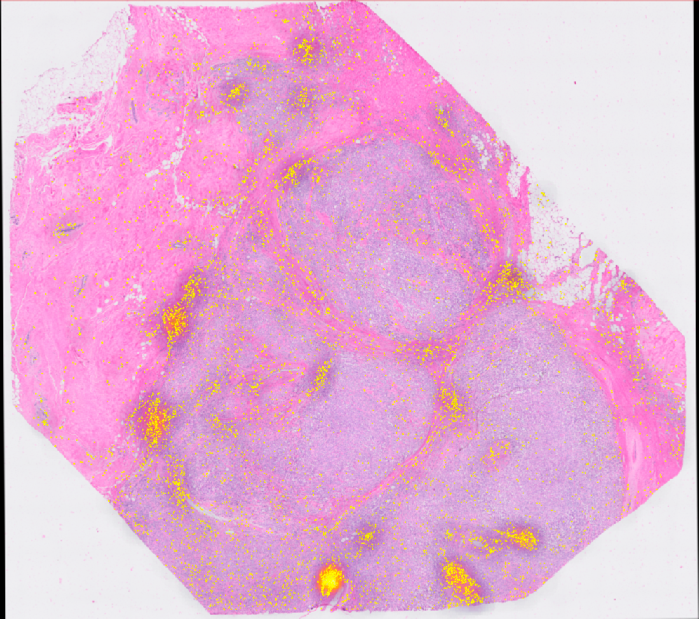}}%
	\hfill%
	\subfloat[The corresponding ROC/AUC.]{\label{fig:result4:d}\includegraphics[width=0.33\linewidth]{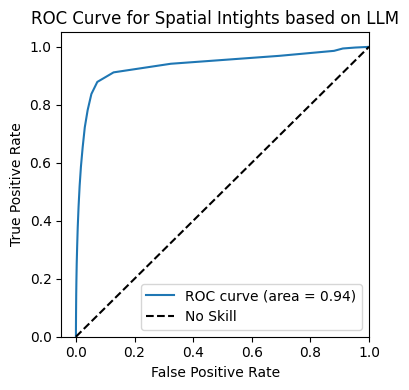}}%
\caption{The result showcases the discovery of spatial insights based on human languages using a LLM. (a) is the prompt used in this experiment, which is identical to the LLM-based interpretation showing in Figure~\ref{fig:result1:c}. (b) shows the the density map of the ground truth indicated by yellow spots. (c) illustrates the density map of the predicted outcome. (d) presents the ROC/AUC curve, indicating the accuracy of the prediction based on the user's description provided in human language.}
	\label{fig:result4}
\end{figure}

In this sub-section, we will present using LLM to identify single-cell clusters or regions that reflects the description of the given prompt. The prompt used was:
\begin{center}
	\parbox{0.9\textwidth}{
		\texttt{``Identify corresponding gene ontology term IDs from below content, and return the result in json format. \\\\\{PROMPT\},''}}%
\end{center}
where the PROMPT was given by the user describing the desired biological status. 

Figure~\ref{fig:result4} demonstrates the experiment using the same prompt as Figure~\ref{fig:result1:c} to compare forward and backward analyses outcomes. We identified single-cell clusters via gene expression levels and used the ROC method to gauge model accuracy, yielding a high-quality prediction with an AUC of 0.94, as shown in Figure~\ref{fig:result4:d}. The predicted outcome confirmed the role of immune response in T cell activation, vasculature development, and granzyme production, showcasing QuST-LLM's ability to reveal spatial insights using human languages.

\section{Conclusion}

The integration of large language models (LLMs) into spatial transcriptomics (ST) analysis, embodied in the QuST-LLM tool, signifies a considerable leap in the genomics field. QuST-LLM's ability to transform intricate, high-dimensional ST data into comprehensible biological narratives significantly enhances the interpretability and accessibility of ST data. Our study has demonstrated that QuST-LLM is not only capable of interpreting biological spatial patterns, but it can also identify specific single-cell clusters or regions based on user-provided natural language descriptions. This represents the transformative potential of LLMs in computational biology research and their ability to assist researchers in deciphering the spatial and functional complexities of tissues.

One avenue for future research could be the fine-tuning of the underlying LLM. By training the model on more specific biological and genomic datasets, it could further improve the tool's ability to interpret and analyze ST data. This could lead to more precise interpretations and potentially uncover deeper insights, thereby driving further advancements in biomedical research.

Overall, QuST-LLM represents a significant step forward in making ST data analysis more intuitive and accessible. With the prospect of fine-tuning the LLM, we anticipate even more impactful contributions to the field in the future.

\section{Availability}

The QuST-LLM is is a function provided in QuST, which is developed based on QuPath 0.5.1 and Python 3.10+ and is available under the Apache 2.0 license (\url{https://github.com/huangch/qust}). A user guide is provided at \url{https://github.com/huangch/qust/user_guide}, including a step-by-step tutorials, GPU support, and examples demonstrating the use of the extension in analysis pipelines.

\bibliographystyle{unsrtnat}
\bibliography{qustllm}  

\end{document}